\documentclass[JHEP,12pt]{article}

\usepackage{booktabs}

\usepackage{jheppub}
\usepackage{amssymb,amsmath,amsthm}
\usepackage[shortlabels]{enumitem}
\usepackage{braket}
\usepackage{breqn,hyperref}
\usepackage{comment} 
\usepackage{caption}
\usepackage{subcaption}
\usepackage{graphicx, xcolor, varwidth}
\usepackage{color}
\usepackage{float}
\theoremstyle{definition}

\theoremstyle{remark}

\DeclareMathOperator{\tr }{tr}

\let\a=\alpha    
   
     \let\r=\rho

  \let\D=\Delta

\title{Page transition for the complexity of an evaporating black hole}
\author[a]{Violet Concepcion,}
\author[a,b,c,d]{Yasunori Nomura,}
\author[a,e]{Kyle Ritchie,}
\author[a]{Samuel Weiss}
\affiliation[a]{Leinweber Institute for Theoretical Physics and Department of Physics,\\
University of California, Berkeley, CA 94720, USA}
\affiliation[b]{Theoretical Physics Group, Lawrence Berkeley National Laboratory,\\
Berkeley, CA 94720, USA}
\affiliation[c]{RIKEN Center for Interdisciplinary Theoretical and Mathematical Sciences (iTHEMS),\\
RIKEN, Wako 351-0198, Japan}
\affiliation[d]{Kavli Institute for the Physics and Mathematics of the Universe (WPI), UTIAS,\\
The University of Tokyo, Kashiwa, Chiba 277-8583, Japan}
\affiliation[e]{Quantum Algorithms and Applications Collaboratory, Sandia National Laboratories, Livermore, CA, USA}

\emailAdd{violet\_concepcion@berkeley.edu}
\emailAdd{ynomura@berkeley.edu}
\emailAdd{kyle\_ritchie@berkeley.edu}
\emailAdd{sam.weiss@berkeley.edu}
\makeatletter
\gdef\@fpheader{\mbox{}}
\makeatother

\abstract{Recent results~\cite{Fan:2025moc,Haah:2025hyf} demonstrate that there exists a sharp, Page-like transition for the complexity of subsystems of Haar-random states as their fractional subsystem size surpasses one half. They further demonstrate that this transition also occurs for the holographic complexity of boundary subregions of eternal AdS black holes, assuming the Complexity=Volume (CV) proposal for subregions. We interpret this transition as a crossover from spectrum-dominated to basis-dominated subsystem complexity, reflecting the breakdown of approximate thermality beyond half-system size. We then apply this reasoning to an evaporating AdS black hole coupled to a bath, modeled by a quantum circuit undergoing random evolution on an interior subsystem of diminishing size. Using the basis-spectrum decomposition of subsystem complexity, we argue for a similar Page-like transition in the radiation complexity. We then show, using CV for subregions, that the same transition appears in the holographic complexity of the radiation subsystem through the emergence of an island, whose volume gives the dominant contribution. We argue that the island volume contribution resolves an apparent complexity paradox analogous to the information paradox.}

\begin{document}
\maketitle

\section{Introduction}

In recent decades, there has been enormous progress in our understanding of information flow in gravity. By understanding holographic duality~\cite{Maldacena:1997re,Aharony:1999ti} as an approximately quantum error-correcting map between the gravitational and CFT Hilbert spaces~\cite{Almheiri:2014lwa,Pastawski:2015qua,Harlow:2016vwg}, a precise relationship between entanglement and geometry has been formulated~\cite{Ryu:2006bv,Ryu:2006ef,Hubeny:2007xt,Faulkner:2013ana,Engelhardt:2014gca}. This progress is perhaps best illustrated by developments in entanglement wedge reconstruction~\cite{Jafferis:2015del,Dong:2016eik}, including the derivation of the Page curve~\cite{Page:1993wv} for the entropy of black hole radiation~\cite{Penington:2019npb,Almheiri:2019psf,Almheiri:2019hni,Penington:2019kki,Almheiri:2019qdq}.

Many recent proposals suggest that computational complexity may share a similar relationship to geometry~\cite{Susskind:2014rva,Stanford:2014jda,Brown:2015bva}, though ambiguities between different complexity measures, and lack of a precise mathematical framework for computing complexity in quantum field theories have prevented complexity from becoming a precise entry in the holographic dictionary~\cite{Jefferson_2017,Belin:2021bga,Belin:2022xmt,Myers:2024vve}.

This is especially true in the context of dynamical gravitational systems, such as an evaporating black hole. A naive application of one of the most robust putative duals, the ``Complexity = Volume'' (CV) proposal~\cite{Susskind:2014rva,Stanford:2014jda,Susskind:2014moa}, leads to an apparent contradiction. For example, after the black hole has finished evaporating, the bulk spacetime no longer has a black hole. Thus, naively, the vacuum subtracted volume of the spacetime is approximately zero, and if $C=V_{\rm BH}$ holds, it implies that the complexity of the radiation subsystem vanishes. This would suggest that the radiation is exactly thermal, which is analogous to the naive semiclassical result that the entanglement entropy of the radiation grows without bound~\cite{Hawking:1976ra}. 

\begin{figure}[t]
\centering
  \begin{subfigure}[b]{0.45\textwidth}
  \includegraphics[width=\linewidth]{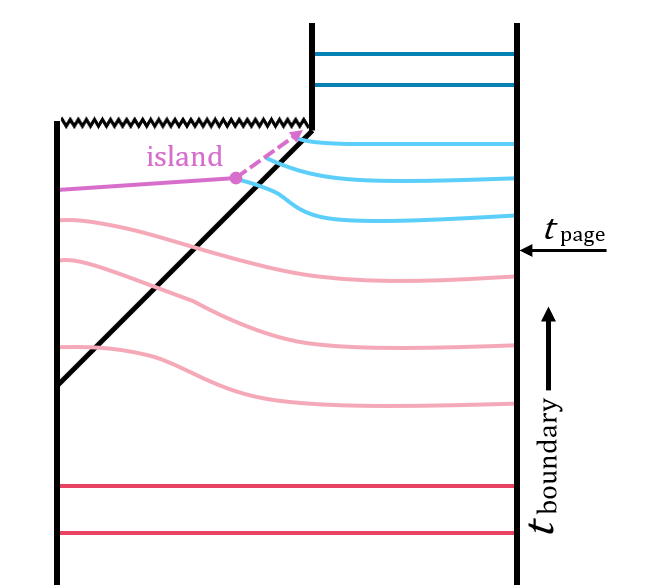}
  \caption{}
  \end{subfigure}
  \begin{subfigure}[b]{0.45\textwidth}
  \includegraphics[width=\linewidth]{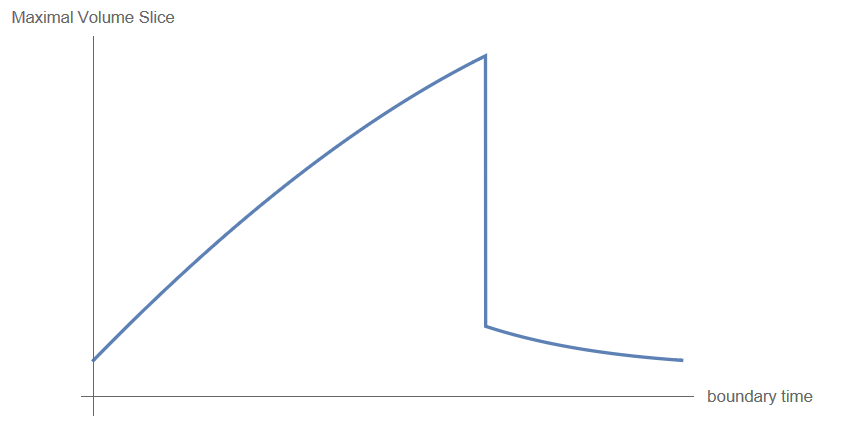}
  \caption{}
  \end{subfigure}
\caption{(a) We sketch the maximal volume slices associated with the boundary at a given time in the AdS/Schwarzschild Penrose diagram. Before the Page time, we have the light red slices which go through the interior of the black hole. The blue slices end on the edge of the island, since they are the maximal volume slices through the entanglement wedge of the CFT living on the boundary, which no longer includes the island. At late times, the dark blue slices represent the post-black hole near-flat maximal volume slices. (b) We depict the value of the volume qualitatively. It increases at a rate proportional to $M^2$ until the time the island forms, at which point the volume drops dramatically.}
\label{fig:CVNaive}
\end{figure}

A more plausible candidate for a holographic complexity dual is the volume of the maximal slice through the entanglement wedge: 
\begin{equation}
  C(\rho_a) = \max\limits_{\Sigma\in e(a)}\text{Vol}(\Sigma)
\end{equation}
where $\rho_a$ is the reduced density matrix of a boundary subregion $a$, and $\Sigma$ is a partial Cauchy slice through its bulk entanglement wedge $e(a)$ \cite{Carmi_2017}. The crucial difference between this proposal and the naive application of $C=V_{\rm BH}$ is the appearance of the entanglement island after the Page time~\cite{Penington:2019npb,Almheiri:2019psf,Almheiri:2019hni}. The island remains even after evaporation is complete, offering a non-vanishing (vacuum subtracted)%
\footnote{Throughout the document, we always take the volume contribution to be vacuum-subtracted so that IR divergences arising from slices extending to the asymptotic boundary may be ignored. This has no effect on the growth rate of the holographic complexity.}
volume contribution, suggesting that the complexity saturates near the maximal volume slice through the black hole interior. On the other hand, before the Page time, the entanglement wedge of the radiation contains only the black hole exterior, yielding an approximately vanishing volume contribution. This suggests that there is a sharp jump in the radiation complexity at the Page time, which seems to indicate that radiation state after the Page time is highly distinguishable from a thermal state of the same size. See Fig.~\ref{fig:CVNaive}.

A similar sharp transition for holographic subsystem complexity was demonstrated in~\cite{Fan:2025moc}. The authors in that work find a jump in the holographic complexity near the Page transition~\cite{Page:1993df} for subregions of varying size of the AdS boundary, in static spacetimes with and without black holes. They find that when the boundary subregion surpasses half of the total system size, the entanglement wedge (and hence its volume) undergoes a sharp transition. In the two sided black hole geometry, for example, the entanglement wedge of a subsystem consisting of two symmetric disconnected regions on opposite sides of the black hole exterior undergoes a transition between disconnected and connected phases. In the latter phase, the entanglement wedge includes the black hole interior, while the former does not. 

Compellingly, they find a similar transition in the subsystem complexity for states prepared by random circuits in finite-dimensional qubit models, thought to accurately represent the statistics of BH microstates~\cite{Hayden:2007cs,Sekino:2008he,Brown:2017jil}. In particular, they find that for systems with fractional size less than half, the saturation value of the complexity is polynomial in the subsystem size, while for subsystems of size greater than one half, the saturation value is exponential. One expects that the complexity of typical states is approximately saturated, implying a jump in the subsystem complexity.

This raises the question of whether the jump in subsystem complexity found in~\cite{Fan:2025moc} is the same as that expected from the inclusion of the island in the entanglement wedge. Namely, as the radiation subsystem becomes a larger fraction of the total system, is the jump in the holographic complexity/volume reflected in a similar jump in an appropriate circuit model?

To address this question, we will compare the behavior of holographic complexity proposals with several circuit-based notions of mixed-state complexity. Our goal is not to establish an exact duality between a particular complexity measure and a bulk geometric quantity, but rather to determine whether the qualitative features of the holographic transition are reproduced in microscopic models of black-hole evaporation. We will argue that this transition reflects a change from a regime in which the complexity is controlled primarily by the entanglement spectrum to one in which the complexity of the eigenbasis becomes dominant.

An obstacle to straightforwardly applying their argument to the radiation and interior subsystems of an evaporating black hole is that the dynamics of the full system are not accurately modeled by a random circuit. This is because the radiation subsystem does not undergo scrambling dynamics. Therefore, black hole evaporation is more accurately modeled by a system undergoing random circuit evolution on an ``interior'' subsystem of decreasing size, and simple dynamics on the ``exterior'' subsystem whose size is correspondingly increasing. The emission of Hawking quanta corresponds to a reduction in size of the interior subsystem and a compensating increase in the radiation subsystem. The rate of change of subsystem size corresponds to the rate of Hawking emission, which is likewise dependent on the size of the interior. We will develop such a model, illustrating our points.

The organization of this paper is as follows. In Sections~\ref{sec:def} and \ref{sec:setup}, we review the relevant notions of complexity for pure and mixed states, and we detail the setup in which we perform our analysis. In Section~\ref{sec:fullsystem}, we analyze the total system complexity of an evaporating black hole using both holographic arguments, as well as the more realistic model for black hole evaporation described above. In Section~\ref{sec:subsystem}, we analyze the holographic complexity of the black hole and radiation subsystems, showing that the latter exhibits a jump at the Page time due to the appearance of the island. We then demonstrate that the same jump occurs for the radiation subsystem of our qubit model. In Section~\ref{sec:discus}, we conclude by offering an interpretation of this jump in terms of distinguishability from thermality for subsystems whose size exceeds half of the total system, along the lines of Page's original argument~\cite{Page:1993df}, pointing towards a ``Page curve for complexity.'' We argue that, just as the quantum extremal surface (QES) prescription~\cite{Engelhardt:2014gca,Penington:2019npb,Almheiri:2019psf,Almheiri:2019hni} restores the correct entanglement entropy beyond semiclassical effective field theory, the island volume contribution to holographic complexity captures fine-grained information about the radiation that is inaccessible within effective field theory~\cite{Hawking:1976ra}.

While this paper was being prepared, related work~\cite{Bragagnolo:2026ltv} appeared. The authors study an evaporating black hole in JT gravity with an end-of-the-world brane. They compute a boundary-to-brane two-point function and interpret it as a measure of black hole subsystem complexity. They find that it grows until the Page time and falls exponentially afterwards. This behavior agrees qualitatively with the results of~\cite{Fan:2025moc,Haah:2025hyf}, as well as our own. Other recent works on holographic and subsystem complexities include~\cite{Caputa:2026ldd,Fang:2026dte,Concepcion:2026fhv}.

\section{Mixed state complexity}
\label{sec:def}

Below we define the notions of complexity which we use for pure and mixed states, and which we apply to our models of an evaporating black hole. We then argue for a bound on  the complexity of subsystems which are less than half the total system size.

\subsection{Definitions}

In systems with a finite Hilbert space, the complexity of a pure state $\psi\in\mathcal{H}$ is defined as the minimum number of operations needed to prepare $\psi$ from some reference state $\psi_{\rm ref} \in \mathcal{H}$ by a unitary quantum circuit $U:\psi_{\rm ref} \rightarrow \psi$~\cite{Nielsen:2005mn,Nielsen:2006qcg,Dowling:2006tnk}. Formally, we choose with some set of universal gates $\mathcal{G} = \{g_i\}$ and a tolerance $\varepsilon$, and construct a unitary circuit  $U = g_{i_1}\cdots g_{i_N}$ such that $|\langle \psi|U|\psi_{\text{ref}}\rangle|^2>1-\varepsilon$. In the discussion below we will generally suppress the choice of gate set and tolerance. Defining $V_U$ to be the number of gates in circuit $U$, the complexity is: 
\begin{equation}
  C(\psi) = \min\limits_{U:\psi_{\rm ref}\rightarrow\psi}V_U.
\end{equation}

In \cite{Agon:2018zso}, a natural generalization to mixed states was proposed. The {\it purification complexity} $C_P$ of a mixed state $\rho \in \text{End}(\mathcal{H})$ is defined by the minimum number of gates to prepare any purification $\psi$ of $\rho$, starting from a chosen reference $\psi_{\rm ref}$ on $\mathcal{H}$ together with any number of ancilla qubits initialized at $|0\rangle$. That is,
\begin{equation}
  C_P(\rho) = \min_{\substack{n,\psi \\ \psi\text{ purifies }\rho}}\left(\min_{U_n:\psi_{\text{ref}}\otimes |0\rangle^{\otimes n} \to \psi}V_{U_n}\right),
\label{eq:C_P}
\end{equation}
where gates comprising $U_n$ are chosen from the set ${\cal G}_n$, acting on the direct-product space of $\mathcal{H}$ and $n$ qubits. This kind of definition is necessary because it is not possible to change the spectrum of the density matrix by acting with unitaries on the system alone.

It will be useful to consider separately the complexity of producing just the entanglement spectrum of the target sate and the complexity of bring the state obtained in that way to the actual target state. The {\it spectrum complexity} $C_S$ of $\rho$ is defined as the minimal purification complexity of any density matrix with the same spectrum:
\begin{equation}
  C_S(\rho) = \min_{U}C_P(U^\dagger \rho U),
\label{eq:C_S}
\end{equation}
where $U$ is an {\it arbitrary} unitary acting on the Hilbert space $\mathcal{H}$. Since reproducing only the spectrum is weaker condition than reproducing the actual state, $C_S$ is upper bounded by $C_P$: $C_S \leq C_P$.

Similarly, the {\it basis complexity} measures the difficulty of obtaining $\rho$ by acting only with unitaries on $\mathcal{H}$, starting from a reference state $\rho_{\text{spec}}$ with the same spectrum. Defining a reference $\rho_{\text{spec}}$ as the minimal spectrum complexity state with $\rho$'s spectrum, i.e.\ $U^\dagger \rho U$ in (\ref{eq:C_S}) after the minimization, we have%
\footnote{Note that $C_B$ as defined here corresponds to $\tilde C_B$ in~\cite{Agon:2018zso}.}
\begin{equation}
  C_B(\rho) = \min_{U: \rho_{\text{spec}}\to\rho}V_U.
\label{eq:U_B}
\end{equation}
The purification complexity is upper bounded by the sum of $C_S$ and $C_B$, since preparing first the spectrum and then the basis certainly gives a purification of $\rho$. Together with $C_S \leq C_P$, we thus find
\begin{equation}
  C_S(\rho)\leq C_P(\rho) \leq C_S(\rho) + C_B(\rho).
\label{eq:CSbounds}
\end{equation}
In addition, if $\rho$ is obtained by tracing out a subsystem of a pure state $\psi$ in a larger Hilbert space, then by definition
\begin{equation}
  C_P(\rho) \leq C(\psi),
\end{equation}
assuming that the gate set ${\cal G}_n$ used in computing $C_P$ with ${\cal G}$ used in $C$ when the relevant Hilbert spaces coincide.

These observations suggest a qualitative distinction between subsystems smaller and larger than half the total system size, which will be made more precise below. In particular, for a small subsystem of a Haar-random pure state, the density matrix is exponentially close to being maximally mixed~\cite{Page:1993df}. For a maximally mixed density matrix, the basis complexity vanishes, so by (\ref{eq:CSbounds}) $C_P = C_S$. Furthermore, $C_S$ in this case scales linearly with the number of qubits in the subsystem (depending on choice of gate set and reference). By contrast, for subsystems larger than half the total system, one expects the purification complexity to be controlled primarily by the complexity of the underlying pure state, and therefore by the basis complexity. In this regime the basis complexity can be quite large; for example, preparing a generic eigenbasis requires a unitary whose complexity is at worst $\mathcal{O}((\text{dim}\ \mathcal{H})^2)$ gates. A precise result along these lines was given in~\cite{Fan:2025moc}, for states in a one-dimensional system of qubits with periodic boundary conditions, prepared by random quantum circuits.

\subsection{Bounding the complexity of small subsystems}

We now show that the mixed state obtained by tracing out the larger subsystem of a Haar-random pure state indeed vanishing basis complexity, and hence a purification complexity controlled by its spectrum complexity.

From Page's Argument in~\cite{Page:1993df}, we learn that for the two subsystems $A$ and $B$ of a Haar-random pure state whose dimensions $n$ and $m$, respectively, are such that $1\ll m\leq n$, the entanglement entropy between the two subsystems is
\begin{equation}
  S_{m,n} \simeq \ln m -\frac{m}{2n},
\end{equation}
so we see that for the subsystem $B$ of dimension $m$, its state is close to maximally mixed. The deviation from being maximally mixed, or its information, is
\begin{equation}
  \D S = s_B - S_B = \frac{m}{2n} + O\left(\frac1{mn}\right),
\end{equation}
where $s_B = \ln m$ and $S_B$ are the maximum entropy and von~Neumann entropy of $B$, respectively, so $\D S$ is always less than one half of a natural logarithmic unit.

We now discuss the complexity that the state $\r_B$ of this subsystem might have. Let us first discuss $C_B(\rho_B)$. We want to consider the size of space of states close to being maximally mixed. Let us choose our reference state to be, without loss of much generality, the maximally mixed state $\mathbb{I}/m$. We can define the relative entropy to be
\begin{equation}
  D\left(\r_B \middle\| \frac{\mathbb{I}}{m} \right) = \text{Tr} \left(\r_B \ln \r_B - \r_B \ln\frac{\mathbb{I}}{m}\right) =\ln m - S(\r_B) = \D S.
\end{equation}
We can then use Pinsker's Inequality to say that (chapter~13.2 of~\cite{Bengtsson:2017gqs})
\begin{equation}
  D\left(\r_B \middle\| \frac{\mathbb{I}}{m} \right) \geq \frac 1 2 \left\| \r_B - \frac I m \right\|_1^2.
\end{equation}
Here we have used the 1-norm $\left\| A \right\| = \tr \sqrt{A^\dagger A}$. Therefore, we have
\begin{equation}
 \left\| \r_B - \frac{\mathbb{I}}{m} \right\|_1 \leq \sqrt{2\D S} \simeq\sqrt{\frac m n},
\end{equation}
so the smaller subsystem lies very close to being maximally mixed, by a distance scaling with the square root of the ratio of the sizes of the subsystems. If the subsystems $A,B$ consist of $M,N$ qubits respectively, then $m=2^M,n=2^N$. In this case,
\begin{equation}
 \left\| \r_B - \frac{\mathbb{I}}{m} \right\|_1 \lesssim 2^{-\frac{N-M}{2}}.
\end{equation}
This is the precise sense in which $\rho_B$ is close to being maximally mixed.\\

Now we use this to bound the basis complexity of $\rho_B$. Let $\mathcal{R}(\rho_B)=\{U^\dagger \rho_B U\ |\  U\in U(m)\}$ be the space of mixed states with the same spectrum as $\rho_B$. Recall that $C_B(\rho_B)$ is the minimal number of gates in a unitary circuit $U$ on $B$, such that $\left\| U^\dagger \rho_B U - \rho_{\text{spec}} \right\|_1 < \varepsilon$. Here $\rho_{\text{spec}} \in \mathcal{R}(\rho_B)$ is the state with the minimum spectrum complexity. Now, if $\rho_B$ is close to maximally mixed in the sense above, then so is every other state in $\mathcal{R}(\rho_B)$, since
\begin{equation}
  \left\| U^\dagger \rho_B U - \frac{\mathbb{I}}{m} \right\|_1 = \left\| U^\dagger \left(\rho_B - \frac{\mathbb{I}}{m}\right) U \right\|_1 = \left\| \rho_B - \frac{\mathbb{I}}{m} \right\|_1.
\end{equation}
Here, we have used the fact that the 1-norm is invariant under unitary conjugation. Thus, the whole space $\mathcal{R}(\rho_B)$ lies inside a ball of radius $\sqrt{2\Delta S}$. If we choose our tolerance to satisfy
\begin{equation}
  \varepsilon>\sqrt{2\Delta S} = 2^{-\frac{N-M}{2}},
\label{eq:tolerance_bound}
\end{equation}
then immediately $C_B(\rho_B)=0$. For any fixed tolerance $\varepsilon$, this condition is automatically satisfied when $N-M$ is sufficiently large. In this case, the total complexity of $\rho_B$ is equal to its spectrum complexity because of (\ref{eq:CSbounds}).

The spectrum complexity of the system will be bounded by a quantity that scales with the entropy of the system, since it is the number of operations that are needed to purify the state $\r_B$. For a maximally mixed state on $M = \ln m$ qubits, an explicit purification can be prepared using $O(M)$ gates. Conversely, any purification must generate entropy $S = M \ln 2$, and a $k$-local gate can increase the entropy by at most $O(1)$. Therefore, if $\varepsilon \gg \sqrt{m/n}$, then
\begin{equation}
  C_P(\rho_B) = C_S(\rho_B) \sim \ln m.
\end{equation}
We will use this result later when we discuss expectations on the complexity of the radiation state before the Page time, and the complexity of the AdS state after the Page time.

\section{CV for an Evaporating Black Hole}
\label{sec:setup}

In this section, we introduce the setup we are considering: an evaporating black hole in AdS coupled to a bath. We then briefly review the QES prescription and CV for subregions, and outline its extension to include islands. Then, we provide a thought experiment that justifies why the naive CV is not sufficient and must be extended. This will lay the groundwork for a connection to the black hole information paradox.

\subsection{Our Setup:\ A Black Hole in AdS Coupled to a Bath}

We consider the following setup: AdS$_d$ spacetime coupled to a separate bath system $\mathcal{R}$~\cite{Penington:2019npb,Almheiri:2019psf,Almheiri:2019hni}. At some boundary time $t=0$ we insert spherically symmetric null operators that collapse into the bulk to create a black hole.

We can treat this initial state as our reference state. After the black hole forms, it emits Hawking radiation~\cite{Hawking:1974sw}, which is absorbed into the bath and not reabsorbed by the black hole. The coupling of AdS$_d$ with the bath system prevents the black hole from coming into equilibrium and causes it to fully evaporate.

\begin{figure}[t]
\centering
  \includegraphics[width=8cm]{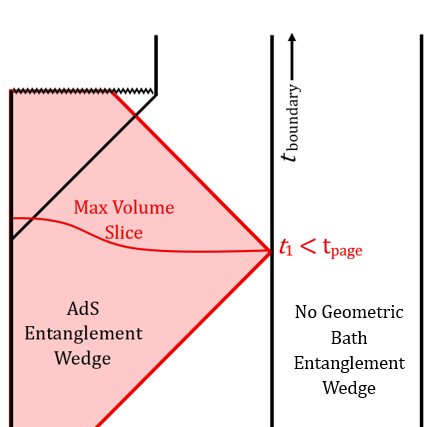}
\caption{Before the Page time $t_{\rm Page}$, the entanglement wedge $e(\mathcal{R})$ of the bath system is the bath itself. The entanglement wedge $e(\mathcal{B})$ of the CFT (living on the AdS boundary) at time $t_1 < t_{\rm Page}$ is the region in AdS$_d$ that can be reached from the equal-time hypersurface of the boundary at $t_1$.}
\label{fig:ewbefore}
\end{figure}
Before the Page time $t_{\rm Page}$, the entanglement entropy of the non-gravitating bath system $\mathcal{R}$ is simply $\mathcal{R}$ itself:\ $e(\mathcal{R})=\mathcal{R}$. In particular, its entanglement wedge has no bulk component. The entanglement wedge $e(\mathcal{B)}$ of the asymptotic boundary $\mathcal{B}$ of AdS$_d$, excluding the bath, is the entire AdS$_d$ region that can be reached from the boundary at that time. The entanglement wedges $e(\mathcal{R})$ and $e(\mathcal{B})$ at a time $t_1 < t_{\rm Page}$ are depicted in Fig~\ref{fig:ewbefore}.

\begin{figure}[t]
\centering
  \includegraphics[width=10cm]{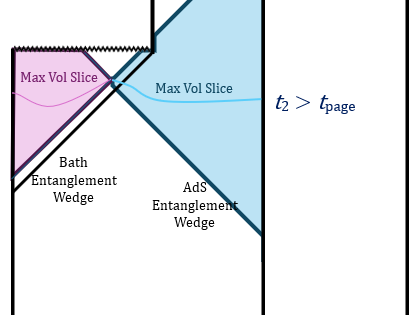}
\caption{After the Page time, the bath's entanglement wedge (pink) now includes the black hole interior. The entanglement wedge of the CFT degrees of freedom (blue) extends to the black hole exterior.}
\label{fig:ewafter}
\end{figure}
After the Page time, the entanglement wedge of the radiation includes the island region, which overlaps with the black hole interior. The entanglement wedge of the CFT that lives on the asymptotic boundary and does not include the non-gravitating bath now contains the black hole exterior; see Fig.~\ref{fig:ewafter}

\subsection{CV for Subregions}

CV for subregions~\cite{Carmi_2017} proposes that the holographic complexity of a boundary subregion $a$ with reduced density matrix $\rho_a$ is equal to the volume of the maximal volume slice $\Sigma$ through its bulk entanglement wedge $e(a)$:
\begin{equation}
  C(\rho_a) = \max_{\Sigma\in e(a)}\text{Vol}(\Sigma).
\end{equation}
The original application of this idea is in the context of using the HRT prescription~\cite{Hubeny:2007xt} to determine the wedge, without including bulk quantum corrections. While it has been compellingly argued that this prescription works for the AdS vacuum and for eternal AdS black holes, it does not correctly capture the holographic complexity of dynamical black holes; see Fig.~\ref{fig:CVNaive}.

This is not entirely unexpected. This prescription of CV for subregions is analogous to using the HRT prescription for calculating the von~Neumann entropy of the subregion $a$. It is well known that the HRT prescription fails when naively applied to the radiation region of evaporating black holes, as it fails to capture the subleading effects due to the bulk effective field theory. These effects are properly captured by the QES formula~\cite{Faulkner:2013ana,Engelhardt:2014gca}
\begin{align}
  S(\rho_a) &= \min\left\{ \operatorname*{ext}_{\gamma}\left( S_{\rm EFT}(\Sigma) + \frac{\text{Area}(\gamma)}{4G_N}\right)\right\}
\nonumber\\
  &= S_{\rm EFT}(e(a)) + \frac{\text{Area}(\partial e(a))}{4G_N},
\end{align}
where $\gamma$ is a codimension-2 surface homologous to the $a$, $S_{\rm EFT}(\Sigma)$ is the entropy of the effective field theory (EFT) degrees of freedom on $\Sigma$ in the bulk, and $e(a)$ is the entanglement wedge of $a$.

When applied to an evaporating AdS black hole in which the Hawking radiation is collected in a non-gravitating bath region $\mathcal{R}$ coupled to the CFT, the disconnected island $\mathcal{I}$ may appear in the entanglement wedge of the radiation, and contributes to the boundary entanglement entropy 
via~\cite{Penington:2019npb,Almheiri:2019psf,Almheiri:2019hni}
\begin{align}
  S(\rho_{\mathcal{R}}) &= \min\left\{\operatorname*{ext}_{\mathcal{I}}\left(S_{\rm EFT}(\mathcal{R} \cup \mathcal{I}) + \frac{\text{Area}(\partial\mathcal {I})}{4G_N}\right)\right\}
\nonumber\\
  &= S_{\rm EFT}(e(\mathcal{R})) + \frac{\text{Area}(\partial e(a))}{4 G_N}
\label{eq:QEI}
\end{align}
where $e(\mathcal{R}) = \mathcal{R} \cup \mathcal{I}_{\rm ext}$, and $\mathcal{I}_{\rm ext}$ extremizes the functional in the first line.

We propose an analogous generalization to the CV formula. For a boundary region $a$, we conjecture
\begin{align}
  C_P(\rho_a) = C_{\rm EFT}(e(a)) + \max_{\Sigma\in e(a)}\frac{\text{Vol}{(\Sigma)}}{G_N\ell},
\label{eq:proposal}
\end{align}
where $C_P(\rho_a)$ is the purification complexity of the state $\rho_a$;%
\footnote{As is common in holographic complexity~\cite{Jefferson_2017,Belin:2021bga,Belin:2022xmt,Myers:2024vve}, the left-hand side of (\ref{eq:proposal}) may not exactly be the purification complexity but some quantity that behaves similarly to it. On the other hand, it cannot exhibit the behavior of either the spectrum complexity $C_S$ or the basis complexity $C_B$, since neither reproduces the holographic behavior discussed later; see, e.g., Section~\ref{subsec:sharp}.}
$C_{\rm EFT}(e(a))$ is the complexity of semiclassical degrees of freedom in $e(a)$, and $\Sigma$ is a slice through the bulk component of the entanglement wedge of $a$. Note that since we are extending CV to the case of AdS coupled to a nongravitating bath, we must specify that the slice subtends the \textit{bulk} component of the entanglement wedge. There is no meaningful notion of a ``slice'' through the bath, which may be entirely non-geometric.%
\footnote{
 This is different from the case of the QES formula for entanglement entropy, where we need not specify that we take the \textit{bulk} component of the entanglement wedge, since the geometric contribution to the generalized entropy, $\rm{Area}(\partial e(a))$ is always contained in the bulk, and therefore always has a well-defined area.
}

Applied to the radiation subsystem of an evaporating AdS black hole coupled to a bath, this proposal gives
\begin{equation}
  C_P(\rho_{\mathcal{R}}) = C_{\rm EFT}(\mathcal{R}\cup\mathcal{I}_{\rm ext}) + \max_{\Sigma\in\mathcal{I}_{\rm ext}}\frac{\text{Vol}(\Sigma)}{G_N\ell}.
\label{eq:CV-Island}
\end{equation}
There is an important subtlety in this expression, which occurs when no island is present. In the case where there is (pure) matter in the bath region which is unentangled with the bulk, the QES formula (\ref{eq:QEI}) simply gives $\mathcal{I}=\emptyset$, so that $e(\mathcal{R}) = \mathcal{R}$ and $S(\rho_{\mathcal{R}}) = S_{\rm EFT}(\mathcal{R}) = 0$. However, the complexity formula (\ref{eq:CV-Island}) does \textit{not} vanish:\ it includes the EFT complexity of the matter in the bath, i.e.\ $C_P(\rho_{\mathcal{R}}) = C_{\rm EFT}(\mathcal{R})$.

To understand this further, we are going to investigate a thought experiment involving a quantum computer without black holes.

\subsection{A Quantum Computer Floating in Space}

Consider empty AdS coupled to a bath, and suppose we introduce a quantum computer in the bulk. The computer is small enough that it does not meaningfully deform the surrounding spacetime. The computer performs computations at a constant rate by applying $p$-gates on sets of $p$ qubits at a time, affecting all of the qubits at each time step, so that the computational complexity of our combined CFT+bath holographic dual system will correspondingly increase. In times far from the saturation value, this growth rate will be approximately linear in time, since switchbacks will be rare if the number of qubits is large, and we assume that it indeed is.
 
The computer uses qubits made of matter which, by some mechanism, slowly evaporate into lightlike radiation which enters the bath in finite time. As the computer evaporates, its computational power will decrease and therefore the rate of complexity growth, will slowly decrease. This radiation will reach the absorbing boundary conditions and join the bath. By the end of the process, we can imagine that the entire computer has completely evaporated, and our system will consist of the AdS vacuum, and a state of propagating free photons in the non-gravitating bath.

Throughout this evaporation process, the naive CV correspondence is unhelpful for measuring the complexity of the computer, since the computer does not meaningfully backreact on the spacetime. In fact, even if the backreaction were not small, we have no reason to believe that as the computer chugs away at its computation, increasing its complexity, the volume of the spacetime will change at all. Moreover, after evaporation is complete, all of the matter has been transferred to the bath, and only pure AdS remains. Since the computer is not entangled with any matter in the AdS bulk, there is no bulk entanglement wedge, and hence no holograpphic complexity. From the perspective of CV, it is as though the computer never existed at all.

Instead, the complexity is solely accounted for by the EFT degrees of freedom in the bath. For example, we can imagine that the complexity can be defined by the number of operations required, after collecting all the photons in the bath, to recreate the initial state of the computer. For this reason, the complexity of the entire system must not consist solely of contributions from the volumes of bulk subregions, but must also include the complexity of the semiclassical matter itself. This motivates the $C_{\rm EFT}(\mathbf{R})$ term in (\ref{eq:CV-Island}).

To motivate the geometric term in (\ref{eq:CV-Island}), we can imagine that a black hole behaves in much the same way as an evaporating quantum computer, with an important caveat:\ Due to the existence of the horizon, the semiclassical radiation entering the bath remains entangled with interior Hawking partners. In the semiclassical picture, which is appropriate to adopt in using (\ref{eq:QEI}) and (\ref{eq:CV-Island}), the radiation is essentially maximally mixed, and consequently $C_{\rm EFT}$ has little or no basis complexity and is therefore small. The complexity of the final-state radiation in the case of black hole evaporation, thus, is not captured by measures of semiclassical EFT complexity.

This, however, seems reasonable, since the evaporation of a black hole occurs through quantum gravitational effects. We expect that the complexity of Hawking radiation is accounted for, instead, by effects beyond the semiclassical physics, similar to those responsible for the Page curve, i.e.\ the appearance of the island in the QES prescription. While we leave a detailed analysis of such complexity to future work, we are motivated by the success of CV for static spacetimes to propose that this contribution to the complexity appears as the volume of the island. Indeed, absent the volume contribution, we arrive at the unlikely conclusion that the complexity of the final evaporation state is small, even after having undergone the complex dynamics of black hole formation and evaporation.

In the following sections, we apply the proposal (\ref{eq:CV-Island}) to an evaporating AdS black hole coupled to a non-gravitating bath. We show that the holographic complexity of the full system, as well as the radiation subsystem, scales in a manner consistent with what is expected of a quantum circuit undergoing random evolution in a subsystem of decreasing size.

\section{Holographic Complexity of the Full System}
\label{sec:fullsystem}

In this section we apply CV~\cite{Susskind:2014rva,Stanford:2014jda,Susskind:2014moa} to the state of the full boundary system. For the full boundary state, the generalized prescription reduces to the usual CV proposal. We expect that the complexity of the full system is equal to the vacuum subtracted volume of the maximal slice through the entire interior geometry. The portion of the slice through the exterior contributes approximately zero vacuum subtracted volume, so that the dominant contribution arises from the volume of the maximal slice through the interior.

We first discuss the interior volume of an evaporating black hole. We find that after transient effects associated with black-hole formation, the volume of the maximal slice grows linearly in time. Because the AdS system is coupled to an external bath through absorbing boundary conditions, the black hole will continue to evaporate. The volume of the maximal slice through AdS will grow at a rate proportional to the remaining mass. We then compare this to a circuit model for an evaporating black hole, whose full system complexity growth mirrors the growth of the black hole interior.

\subsection{The Eternal Black Hole in AdS}

To warm up, we calculate the volume of the maximal slice through the interior of an eternal black hole in $D$-dimensional AdS spacetime. We review the controlled computation in the static case following~\cite{Stanford:2014jda,Susskind:2014moa}.

The metric is given by
\begin{equation}
  ds^2 = -f(r) dt^2 + \frac{dr^2}{f(r)} + r^2 d\Sigma^2_{D-2}, 
\qquad
  f(r_h)=0.
\end{equation}
Passing to ingoing Eddington--Finkelstein coordinates,
\begin{equation}
  ds^2 = -f(r) dv^2 + 2 dv dr + r^2 d\Sigma^2_{D-2}.
\end{equation}
The volume functional for a codimension-1 slice $v(r)$ is
\begin{equation}
  V = \Omega_{D-2} \int\! dr\, r^{D-2} \sqrt{2 v'(r) - f(r) v'(r)^2}.
\label{eq:Vol}
\end{equation}

Since the volume functional does not depend explicitly on $v$, the corresponding conjugate momentum is conserved:
\begin{equation}
  E = \frac{r^{D-2} (1 - f v')}{\sqrt{2 v' - f {v'}^2}}.
\end{equation}
Solving for $v'(r)$ then gives
\begin{equation}
  v'(r) = \frac{1}{f(r)} \left(1 - \frac{E}{\sqrt{E^2 + r^{2D-4} f(r)}} \right).
\end{equation}
Plugging this into (\ref{eq:Vol}), the on-shell volume becomes
\begin{equation}
  V = \Omega_{D-2} \int\! dr\, \frac{r^{2D-4}}{\sqrt{E^2 + r^{2D-4} f(r)}}.
\label{eq:stat-vol}
\end{equation}

At late times the maximal slice spends most of its volume near the radius that maximizes
\begin{equation}
  \mathcal{V}(r) = r^{D-2} \sqrt{|f(r)|}.
\end{equation}
This gives
\begin{equation}
  \left.\frac{d}{dr} \left( r^{D-2} \sqrt{|f(r)|} \right)\right|_{r=r_m} = 0.
\end{equation}
The late-time behavior is then
\begin{equation}
  V(t) \sim v_D\, t,
\qquad
  v_D = \Omega_{D-2} r_m^{D-2} \sqrt{|f(r_m)|}.
\label{eq:stat-growth}
\end{equation}
In the large-AdS-black-hole limit $r_h \gg L$, i.e.\ $M \gg L^{D-3}/G_N$ (where $L$ is the AdS radius), $r_m \sim (G_N L^2 M)^{1/(D-1)}$ and
\begin{equation}
  v_D \sim G_N L M \propto M.
\end{equation}

\subsection{The Evaporating Black Hole in AdS}

We now generalize to the evaporating case. We model the geometry by the Vaidya-like metric
\begin{equation}
  ds^2 = -F(r,v) dv^2 + 2 dv dr + r^2 d\Sigma^2_{D-2},
\end{equation}
with
\begin{equation}
  F(r,v) = \frac{r^2}{L^2} + 1 - \frac{\mu(v)}{r^{D-3}},
\end{equation}
where $\mu(v)$ encodes the time-dependent mass parameter.

We parameterize the slice by $\lambda$:
\begin{equation}
  V = \Omega_{D-2} \int\! d\lambda\, r^{D-2} \sqrt{-F \dot{v}^2 + 2 \dot{v} \dot{r}}.
\end{equation}
Using reparameterization invariance, we choose the gauge%
\footnote{This equation implies that $\lambda$ has dimensions of [length]$^{-D+3}$.}
\begin{equation}
  \sqrt{-F \dot{v}^2 + 2 \dot{v} \dot{r}} = r^{D-2}.
\label{eq:gauge}
\end{equation}
Then the volume simplifies to
\begin{equation}
  V = \Omega_{D-2} \int d\lambda \, r^{2D-4}.
\end{equation}

Define the momentum conjugate to $v$:
\begin{equation}
  P = \dot{r} - F(r,v)\dot{v}.
\label{eq:P-def}
\end{equation}
The resulting first-order equations are
\begin{align}
  \dot{r} &= \pm \sqrt{P^2 + F(r,v) r^{2D-4}},
\label{eq:dot-r}\\
  \dot{v} &= \frac{\dot{r} - P}{F(r,v)},
\\
  \dot{P} &= -\frac{1}{2} \partial_v F \, \dot{v}^2,
\end{align}
where the dot denotes differentiation with respect to $\lambda$. Unlike the static case, $P$ is not conserved.

We now assume
\begin{equation}
  |\partial_v F| \ll |F|/r_h,
\end{equation}
i.e.\ the geometry evolves slowly compared to the local light-crossing time $\sim r_h$. Then, over short segments of the slice, $F(r,v)$ can be treated as approximately static. In this regime, the solution locally resembles the static one, with slowly varying parameters:
\begin{equation}
  r_m \to r_m(v),
\qquad
  v_D \to v_D(v).
\end{equation}
Thus, by applying the static result locally along the slowly evolving geometry, the volume growth is given by
\begin{equation}
  \frac{dV}{dv} = v_D(v) = \Omega_{D-2}\, r_m(v)^{D-2} \sqrt{|F(r_m(v),v)|},
\end{equation}
which reduces to the static expression (\ref{eq:stat-growth}) when $F$ becomes time independent.

For a large AdS black hole $r_h \gg L$, $r_h^{D-1} \approx G_N L^2 M$, and one finds $r_m \approx 2^{-\frac{1}{D-1}}r_h$ and
\begin{equation}
  F(r_m(v),v) \approx -\frac{r_m(v)^2}{L^2}.
\end{equation}
Therefore, in this case $v_D \propto r_h^{D-1}/L$, or
\begin{equation}
  v_D(v) = \alpha\, G_N L M(v),
\label{eq:alpha}
\end{equation}
where $\alpha$ is a $D$-dependent $O(1)$ constant. Then, we have
\begin{equation}
  V(v) = \alpha\, G_N L \int^{v}\! dv' \, M(v'),
\end{equation}
and the volume is controlled by the integrated mass history of the black hole.

The $v$ dependence of the mass can be obtained from the Stefan-Boltzmann law:
\begin{equation}
  \frac{dM}{dv} \propto - r_h^{D-2} T^D.
\end{equation}
With the temperature $T \propto r_h/L^2$ for a large black hole, we can write
\begin{equation}
  \frac{dM}{dv} =  - k \frac{G_N^2 M^2}{L^{2D-4}},
\end{equation}
where the constant $k$ here depends on the strength of coupling between the gravitational system and the bath. Integrating this, we find that the black hole mass decreases as
\begin{equation}
  M(v) = \frac{M_0}{1 + k G_N^2 L^{4-2D} M_0 v},
\end{equation}
and consequently the volume goes as
\begin{equation}
  V(v) = V_0 + \frac{\alpha}{k} \frac{L^{2D-3}}{G_N} \ln\frac{M_0}{M}.
\label{eq:volume}
\end{equation}

The quasistatic approximation is only valid so long as the light crossing time $t_{\rm lc} \sim r_h$ is much smaller than the time for the black hole mass, and hence its entropy, to change by an order-one fraction, the Page time $t_{\text{Page}} \sim L^{2D-2}/(k G_N r_h^{D-1})$. This requires that
\begin{equation}
  k \ll \frac{L^{2D-2}}{G_N r_h^D}.
\label{eq:k-cond}
\end{equation}
This bound will continue to hold if it holds initially. The large black hole approximation will of course not hold throughout evaporation, but will be reliable at least up to the Page time.

\subsection{A Circuit Model for Black Hole Evaporation}

Here we provide a circuit model for black hole evaporation, which consists of $p$-gates acting randomly on subsystems of time-varying size $N(t)$. We first review the time-independent case, and then account for time dependence.

\subsubsection*{Complexity of Random Circuits}

We consider a system of $N$ qubits on which random $p$-local gates act at each time step $\D t$. The resulting random circuit defines a stochastic time evolution whose complexity growth we now analyze.

\begin{figure}[t]
\centering
  \includegraphics[width=7cm]{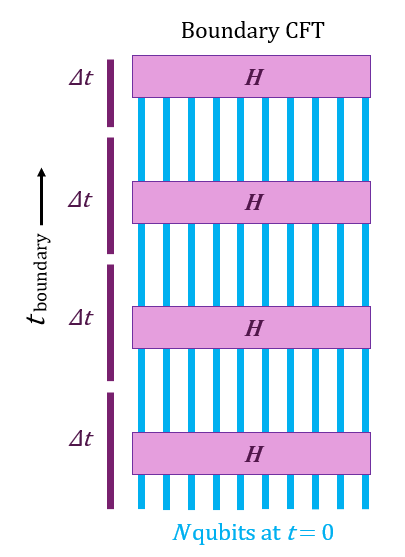}
\caption{A time dependent quantum circuit as a set of $N$ qubits. Every $\D t$ we act on the circuit with an operator $H$ which consists of a product of $p$-gates, acting nonlocally.}
\label{fig:circuitnoevap}
\end{figure}

We can describe this evolution using the Nielsen geometry~\cite{Nielsen:2005mn,Nielsen:2006qcg,Dowling:2006tnk}, where it will take one step in this geometry at each time step. The circuit evolution corresponds to a trajectory through $SU(2^N)$, starting from the identity operator:
\begin{equation}
  \ket{\Psi(n)} = g_n \ldots g_3 g_2 g_1 \ket{\Psi(0)} = u(n) \ket{\Psi(0)},
\end{equation}
where each $g_i$ denotes the combined operator obtained by multiplying the $p$-gates acting during a single time step. We can define the complexity of the state $\ket{\Psi(n)}$ as the smallest number of operations from our set of gates that can create $u(n)$ by acting on the identity operator $\mathbb{I}$. The evolution is generated by a path-dependent Hamiltonian $H(s)$:
\begin{equation}
  \frac{du}{ds} = -i H(s) u,
\qquad
  u(s) = P e^{-i\int\! H(s')\, ds'},
\label{eq:u-s}
\end{equation}
where $P$ denotes a path-ordered product, and $s$ parameterizes time evolution, and $H(s)$ is the time-dependent Hamiltonian. See Fig.~\ref{fig:circuitnoevap}.

The authors of~\cite{Dowling:2006tnk} have defined a complexity metric by
\begin{equation}
  \langle Z_A \cdot Z_B \rangle = \frac{1}{2^N} {\rm Tr}[Z_A (\mathcal{P}+q\mathcal{Q}) Z_B],
\end{equation}
where $Z_A$ and $Z_B$ are $4^N-1$ traceless operators that can be built out of Pauli matrices and identities acting on each qubit. Here, $\mathcal{P}$ projects onto operators acting on at most $p$ qubits, $\mathcal{Q}$ onto operators acting on more than $p$ qubits, and $q \gg 1$ is a cost parameter that penalizes directions generated by operators acting on more than $p$ qubits. With this metric, the evolution generated by an easy Hamiltonian (built only from $p$-local gates) follows a shortest geodesic up to an exponentially long time $s_c \sim e^N$, after which it remains a geodesic but is no longer the shortest one.

In Susskind's epidemic model~\cite{Susskind:2018pmk}, the complexity of a state undergoing $p$-local Hamiltonian evolution satisfies
\begin{equation}
  \frac{dC(\tau)}{d\tau} = s(\tau),
\end{equation}
where $s(\tau)$ is the number of qubits to which the ``infection'' (produced by the action of a simple operator at $\tau=0$) has spread, and satisfies
\begin{equation}
  \frac{ds}{d\tau} = \frac{(N-s)s}{N-1} \approx \frac{(N-s)s}{N},
\label{eq:epid}
\end{equation}
where $N$ is the number of qubits in the system. In the black hole context, the circuit time $\tau$ appearing here is the Rindler time, defined as $d\tau = (2\pi/\beta) dt$, where $t$ is the Schwarzschild time. With $N$ constant, (\ref{eq:epid}) is solved by
\begin{equation}
  s(\tau) = N\frac{ e^{\tau-\tau_*}}{1  + e^{\tau-\tau_*}},
\label{eq:epid-res}
\end{equation}
where $\tau_* = \ln N$ is the scrambling time. The spread saturates to $N$ in $\mathcal{O}(1)$ scrambling times, after which the complexity grows linearly in $\tau$.

\begin{figure}[t]
\centering
  \includegraphics[width=10cm]{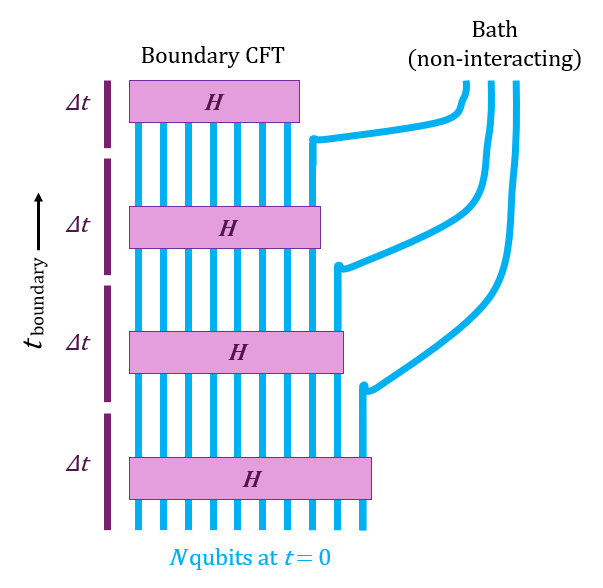}
\caption{A time dependent quantum circuit that more closely resembles a system undergoing evaporation. Since the black hole is a fast scrambler, we see rapid interactions modeled such that an operator $H$ consisting of a product of $p$-gates acts on the circuit in every time interval $\D t$. We also have a bath system in which no interactions occur, and no gates are used. Every $\D t$ we remove one qubit from the circuit to this bath system.}
\label{fig:circuit}
\end{figure}
To crudely model black hole evaporation, we will allow $N$ to depend on time, imagining that qubits are gradually removed from the system to become radiation; see Fig.~\ref{fig:circuit}. The spread should then obey
\begin{equation}
  \frac{ds}{d\tau} = \frac{(N - s)s}{N} + \frac{s}{N}\frac{dN}{d\tau}.
\end{equation}
The second term accounts for already infected qubits being removed:\ we assume that the qubit to be removed is selected uniformly at random. In terms of $u = s/N$, the fraction of remaining qubits infected, this becomes
\begin{equation}
  \frac{du}{d\tau} = u(1-u),
\end{equation}
so the result (\ref{eq:epid-res}) remains valid when $N$ varies with time.

Since we are concerned with effects on the timescale of black hole evaporation, the saturation of the infection is practically instantaneous, and the complexity is well approximated by
\begin{equation}
  C(\tau) = \int\! s(\tau)\, d\tau \sim \int\! N(\tau)\, d\tau.
\end{equation}
Now, taking $N = S_{\text{BH}}$ and using $d\tau = (2\pi/\beta) dt$, we have
\begin{equation}
  C(t) \sim \int_0^t\! S_{\rm BH}(t')\, T_{\rm BH}(t')\, dt',
\end{equation}
implying that the complexity growth follows the Lloyd bound~\cite{Lloyd:2000cry}. For a large AdS black hole, we have $S_{\rm BH} \propto r_h^{D-2}/G_N$, $T_{\rm BH} \propto r_h/L^2$, and $r_h^{D-1} \sim G_N L^2 M$. Using $M(t) = M_0/(1 + k G_N^2 L^{4-2D} M_0 t)$, we obtain
\begin{equation}
  C(t) \propto \int_0^t\! M(t')\, dt' \propto \frac{L^{2D-4}}{G_N^2}\ln \frac{M_0}{M(t)}.
\end{equation}
Comparing this with the expression of the volume (\ref{eq:volume}), we find
\begin{equation}
  C \propto \frac{V}{G_N L},
\end{equation}
up to a $D$-dependent factor.

\section{Holographic Complexity of the Radiation Subsystem}
\label{sec:subsystem}

In this section, we discuss the holographic complexity of the radiation subsystem of an evaporating black hole. We first use $C=\text{Vol}({\rm EW})$ to compute the holographic complexity of the radiation, and find that there is a sharp jump in complexity due to the appearance of the island in the entanglement wedge. We then argue that a similar jump appears for the complexity of the radiation subsystem of our circuit model. We consider the purification complexity of the subsystem, as well as its basis and spectrum complexity, in order to constrain the shape of the complexity curve before and after the Page time. We show that it is consistent with the expectation from $C=\text{Vol}({\rm EW})$.

\subsection{Maximal Volume Slices through Entanglement Wedges}

Before the Page time, we can simply use the result from Section~\ref{sec:fullsystem}, since the maximal volume slice stretches from the boundary to the interior of the black hole. After the Page time, we must perform a new analysis.

In the static case, the volume functional is given by (\ref{eq:stat-vol}). We can define the renormalized volume by subtracting the pure AdS contribution:
\begin{equation}
  \tilde{V} = \lim_{\Lambda\to\infty}
  \left[
    \Omega_{D-2} \int_{r_*}^{\Lambda} dr \, \frac{r^{2D-4}}{\sqrt{E^2 + r^{2D-4} f(r)}} - V_{\text{AdS}}^{\text{vac}}(\Lambda)
  \right],
\end{equation}
where $r_*$ is either $r_h$ or the island radius. Since
\begin{equation}
  f(r) = \frac{r^2}{L^2} + 1 - \frac{\mu}{r^{D-3}},
\end{equation}
the integrand of the first term behaves at large $r$ as
\begin{equation}
  \frac{r^{2D-4}}{\sqrt{E^2 + r^{2D-4} f(r)}} = \frac{L\, r^{D-3}}{\sqrt{1 + \frac{L^2}{r^2}}} \left[1 + O\left( \frac{1}{r^{D-1}} \right) \right].
\end{equation}
This is exactly the same leading behavior as in pure AdS, so the subtraction cancels all divergences.

The remaining finite contribution comes from subleading terms involving $\mu \sim M G_N$. Expanding to a higher order,
\begin{equation}
  \frac{r^{2D-4}}{\sqrt{E^2 + r^{2D-4} f(r)}} = \frac{L\, r^{D-3}}{\sqrt{1 + \frac{L^2}{r^2}}} \left[1 + \frac{\mu L^2}{2 r^{D-1}} + \cdots \right].
\end{equation}
Subtracting the AdS vacuum removes the $1$ term in the parentheses, leaving
\begin{equation}
  \delta V_{\text{outside}} \approx \frac{\Omega_{D-2}}{2} \int_{r_*}^{\infty}\! dr\, \frac{\mu L^3}{r^2 \sqrt{1 + \frac{L^2}{r^2}}}.
\end{equation}
This integral converges and gives
\begin{equation}
  \delta V_{\text{outside}} \approx \frac{\Omega_{D-2}\, \mu L^2}{2} \arcsin\frac{L}{r_*}.
\end{equation}
Thus, the exterior contribution remains finite after vacuum subtraction.

For a large AdS black hole,
\begin{equation}
  r_h \gg L, \qquad \mu \approx \frac{r_h^{D-1}}{L^2}.
\end{equation}
Taking $r_* \sim r_h$, we obtain
\begin{equation}
  \delta V_{\text{outside}} \sim \frac{\Omega_{D-2}\, \mu L^3}{2 r_h}
  \sim \frac{G_N M L^3}{r_h},
\label{eq:after-Page}
\end{equation}
which decreases as $M$ decreases and is proportional to a term that is suppressed in the large black hole limit.

\begin{figure}[t]
\centering
  \begin{subfigure}[b]{0.4\textwidth}
    \includegraphics[width=\linewidth]{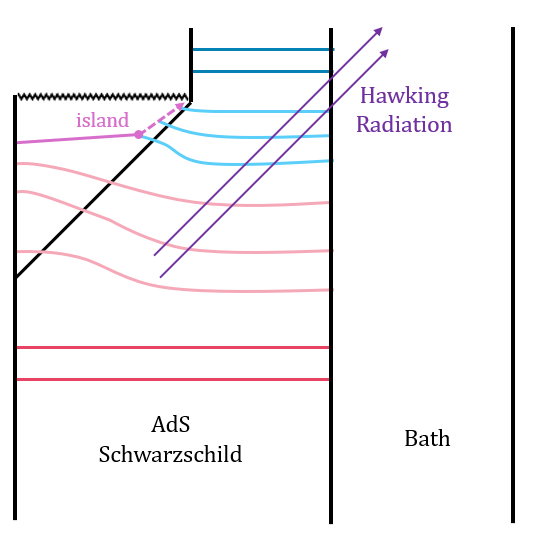}
  \caption{}
  \end{subfigure}
  \begin{subfigure}[b]{0.5\textwidth}
    \includegraphics[width=\linewidth]{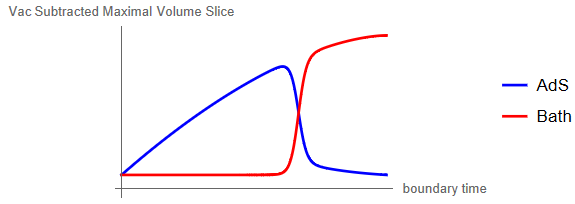}
  \caption{}
  \end{subfigure}
\caption{(a) A sketch of the maximal volume slices in the AdS-Schwarzschild Penrose diagram. We have included a non-gravitating bath into which the Hawking radiation can travel. (b) The qualitative behavior of the volumes associated with the AdS boundary and the bath. The former increases at a rate proportional to $M$ until the time the island forms, at which point it undergoes a sharp transition to a much smaller value. On the other hand, the entanglement wedge of the bath, and hence its complexity, receives a large contribution from the island, undergoing a sharp transition to a much larger value at the Page time.}
\label{fig:CV}
\end{figure}

The result above indicates that the maximal volume of the AdS-boundary entanglement wedge undergoes a sharp transition at the Page time. The ratio of the maximal volumes of the AdS-boundary entanglement wedge immediately after and before the Page time is given by (\ref{eq:after-Page}) and (\ref{eq:volume}) as
\begin{equation}
  \frac{V^{\rm AdS}_{\rm after}}{V^{\rm AdS}_{\rm before}} 
  \sim k \left(\frac{G_N}{L^{D-2}}\right)^2 \frac{M L^2}{r_h} 
  \sim k \frac{G_N r_h^{D-2}}{L^{2D-4}},
\end{equation}
where we have ignored $V_0$ and used $\alpha \approx O(1)$, $M \approx O(M_0)$, and $M \sim r_h^{D-1}/(G_N L^2)$. Using (\ref{eq:k-cond}), we find
\begin{equation}
  \frac{V^{\rm AdS}_{\rm after}}{V^{\rm AdS}_{\rm before}} 
  \ll \left(\frac{L}{r_h}\right)^2
  \ll 1.
\end{equation}
This behavior is depicted in Fig.~\ref{fig:CV}.

Let us now compute the volume of the island appearing after the Page time. In the interior ($r<r_h$), $F(r,v)<0$, and the equation for $\dot{r}$ is given by (\ref{eq:dot-r}). At late times, the solution develops a long segment where $\dot{r} \approx 0$, which occurs when
\begin{equation}
  P^2 + F(r,v)\, r^{2D-4} \approx 0.
\end{equation}
This condition selects the radius $r=r_m(v)$ where
\begin{equation}
  \frac{d}{dr}\left[r^{D-2}\sqrt{|F(r,v)|}\right] = 0.
\end{equation}

Along the $r_m$ segment, we have $\dot{r} \approx 0$, so from (\ref{eq:gauge}), we find
\begin{equation}
  \sqrt{-F(r_m(v),v)}\, |\dot{v}| \approx r_m(v)^{D-2}.
\end{equation}
Thus,
\begin{equation}
  d\lambda = \frac{dv}{|\dot{v}|}
  \approx \frac{\sqrt{|F(r_m(v),v)|}}{r_m(v)^{D-2}}\, dv.
\end{equation}
The volume contribution from the interior is then
\begin{align}
V_{\text{inside}} &= \Omega_{D-2} \int\! d\lambda\, r^{2D-4}
\nonumber\\
  &\approx \Omega_{D-2} \int\! dv\, r_m(v)^{D-2} \sqrt{|F(r_m(v),v)|}.
\end{align}

For a slice anchored at boundary time $t_{\text{bdy}}$ and ending at the island at $v_{\text{is}}$, the interior portion runs from early interior times up to $v_{\text{is}}$:
\begin{equation}
  V_{\text{inside}} = \int^{v_{\text{is}}} dv \, v_D(v),
\end{equation}
where
\begin{equation}
  v_D(v) = \Omega_{D-2}\, r_m(v)^{D-2} \sqrt{|F(r_m(v),v)|}.
\end{equation}
Using the scaling (\ref{eq:alpha}), we obtain
\begin{equation}
  V_{\text{inside}} = \alpha\, G_N L \int^{v_{\text{is}}}\! dv\, M(v).
\end{equation}
At the Page time, this is already large
\begin{equation}
  V_{\text{inside}} \sim \frac{L^{2D-3}}{k G_N} \gg \frac{r_h^D}{L}.
\end{equation}

\subsection{Sharp Transitions in Complexity}
\label{subsec:sharp}

At the Page time, the interior portion of the black hole becomes part of the entanglement wedge of the bath rather than that of the AdS boundary. Therefore, the AdS entanglement wedge will terminate at a short---of order the Planckian---distance inside the black hole horizon. This new boundary near the black hole horizon, i.e.\ the edge of the island, appears at the Page transition and settles to its near-horizon location within a few scrambling times. At late times, when the black hole has completely evaporated, the AdS entanglement wedge will correspond to the vacuum state, which has zero complexity because we subtract the vacuum contribution in our definition of complexity.

At the Page time, the CV proposal predicts a sharp transition in the subsystem complexities. The AdS subsystem loses the largest contribution to its volume, while the bath system now has an entanglement wedge with the same volume.

Does this match the circuit model? To investigate this, we will lean on our two complexity bounds:
\begin{equation}
  C_S(A)\leq C_P(A) \leq C_{B}(A) + C_S(A),
\label{eq:ineq-1}
\end{equation}
\begin{equation}
  C_P(A) \leq C.
\label{eq:ineq-2}
\end{equation}
At early times, the first inequality provides the stronger constraint. This is because the subsystem density matrix looks very close to maximally mixed~\cite{Page:1993df}. By (\ref{eq:tolerance_bound}), a subsystem of $M$ qubits in a system of $M+N$ qubits has zero basis complexity so long as our notion of complexity is defined with a tolerance $\varepsilon > 2^{-\frac{N-M}{2}}$. Except very near the Page time, $N-M$ will be on the order of the initial black hole entropy $S_0$. Thus, the basis complexity of the smaller subsystem will be zero so long as we choose a tolerance $\varepsilon \gg e^{-S_0}$, in other words, so long as we cannot resolve differences in states that are nonperturbative in $G_N$. Under this assumption, we then have $C_P(A) = C_S(A)$, which is of order the entropy of the subsystem.

The relevant inequality will change, however, at the Page time, when the radiation subsystem contains more than half the qubits from the original system and no longer looks maximally mixed. After the Page time, the basis complexity of the radiation subsystem is expected to become huge, scaling polynomially in the dimension of the relevant Hilbert space. This expectation is supported by the fact that the number of gates contributing to the basis complexity is expected to be of the same order as the number of gates needed to decode black hole information in Hawking radiation~\cite{Harlow:2013tf,Hayden:2018khn,Brown:2019rox}. This basis complexity is much larger than the complexity $C$ of the full system, which has been growing linearly in time at a rate proportional to the thermodynamic entropy. Specifically, the complexity of decoding Hawking radiation is expected to scale exponentially with $S$, whereas the complexity of the full system grows only polynomially in $S$. Therefore, after the Page time, the relevant upper bound is the second inequality, Eq.~(\ref{eq:ineq-2}).

This rapid change of the upper bound on the subsystem complexity $C_P$ of the radiation from $C_S$ to $C$ allows the radiation complexity to jump from a small value to a large value around the Page time. This does not, by itself, prove that the radiation complexity actually undergoes such a jump. However, the holographic analysis above predicts precisely such a transition through the appearance of the island. In fact, at the end of evaporation, the radiation subsystem will also be the full system, so we know that at late times the complexity of the radiation subsystem must indeed be close to the complexity of the full system $C$.

This sharp transition is similar to that observed in~\cite{Fan:2025moc}, where both the saturation complexity and the complexity growth rate undergo sharp transitions when the subsystem size crosses one half of the total system. Note, however, that here we are concerned not with the saturation complexity, but with the actual complexity of a subsystem state during the dynamical evolution of the whole system.

\section{Discussion}
\label{sec:discus}

\subsection{A Resolution to the Complexity Information Paradox}

To recap, we first extended the analogy between an AdS black hole and a quantum circuit by including evaporation. We saw that the two models were consistent for the full system, but could only be consistent for subregion complexity when we were careful to include the island in the entanglement wedge of the bath after the Page time.

This picture is consistent with the earlier discussion of an evaporating quantum computer in AdS coupled to a bath. However, rather than computing the complexity of radiation in the bath directly, we lean on the CV correspondence by calculating maximal volume slices in the entanglement wedges of the different subsystems.

Using the CV correspondence is essential for computing the correct complexity, since a naive semiclassical computation incorrectly predicts that the Hawking radiation leaves the black hole in a thermal state~\cite{Hawking:1976ra}. We must include islands to obtain the correct result, much as in the case of entanglement entropy. Just like a generalized entropy, we are led to consider a generalized complexity
\begin{equation}
  C_{\text{gen}} = \frac{V}{G_N L} + C_{\rm EFT},
\label{eq:cgen}
\end{equation}
which corresponds to holographic complexity, a quantity related to the purification complexity for the subsystem.

The highly complex final state of the radiation, generated by the black hole dynamics, can be seen only by correctly implementing CV. Nonperturbative quantum corrections restore the correct behavior by introducing an additional geometric term:\ the volume of the island. This volume term captures the fine-grained complexity of the radiation that becomes accessible when the resolution of the state is taken beyond the limit imposed by effective field theory.

\subsection{Complexity = Action}

The natural Complexity = Action (CA)~\cite{Brown:2015bva,Brown:2015lvg} analogue of this proposal would be to replace our generalized complexity definition \eqref{eq:cgen} with
\begin{equation}
  C_{\text{A,gen}} = \a I_{\rm WDW} + C_{\rm EFT},
\end{equation}
with $I_{\rm WDW}$ the Wheeler-DeWitt action of the entanglement wedge. Conceptually, the same argument fits this setup. Before the Page time, when the radiation subsystem is close to maximally mixed, the CA prescription is dominated by $C_{\rm EFT}$ since there is no island. After the Page transition, the island enters the radiation entanglement wedge, and the action contribution becomes suddenly very large and dominant. Instead of calculating the maximal volume slice through the region, we simply integrate over the region to find the action. In the eternal black hole, $I_{\rm WDW}$ scales with the constant mass. In the case of slow evaporation, $I_{\rm WDW}$ grows at a rate proportional to the dynamic mass as with the maximal volume slice~\cite{Chapman:2018dem}.

We expect that the main place where the story would differ is at the transition. In CA, null joints and corner terms matter in an essential way \cite{Lehner:2016vdi}, and subregion CA can contain area-like pieces that are absent in subregion CV. This would likely lead to a different transition profile but the important points of the story will remain unchanged.

\subsection{Complexity of Reconstruction}

Other authors have investigated the complexity of one-sided reconstruction~\cite{Brown:2019rox}, building on the decoding-complexity perspective of Harlow and Hayden~\cite{Harlow:2013tf}. The complexity of the black hole in this sense is exponential in the system size (i.e., $e^{S_{\rm BH}}$). This is because the authors are looking at the complexity restricted to gates that can only act outside the black hole. Decoding the entire system by only acting on a subsystem is in general a very difficult task, as the authors explain, and so the complexity scales exponentially in the system size. This leads to the inability to reach information inside the black hole semiclassically.

In contrast, the complexity considered in this paper is not the complexity of reconstructing the complementary subsystem from the radiation, but the complexity of the reduced state of the subsystem itself. The relevant gate set is therefore the one used to prepare a purification of $\rho_{\mathcal{R}}$, rather than a decoding circuit whose task is to recover interior operators. The complexity considered here is instead governed by the scrambling dynamics of the black hole degrees of freedom. For an eternal black hole, Susskind et al.~\cite{Susskind:2018pmk} argued that this linear growth saturates a proposed Lloyd bound~\cite{Lloyd:2000cry}. For an evaporating black hole, the growth is sublinear since, over time, less and less of the system is experiencing scrambling (the radiation is taken to the boundary/reservoir). Consequently, this setting is more naturally compared with a simple model of random circuit growth and a random walk through Nielsen geometry.

\section*{Acknowledgments}

We thank Stefan Hollands, Sami Kaya, Hugo Marrochio, Matthew Self, Brian Swingle, and Liz Wildenhain for helpful discussions. This work was supported in part by the U.S. Department of Energy, Office of Science, Office of High Energy Physics under QuantISED award DE-SC0019380 and contract DE-AC02-05CH11231. The work of Y.N. was also supported by MEXT KAKENHI Grant Number JP25K00997.

\bibliographystyle{JHEP}
\bibliography{refs.bib}

\end{document}